# On the Importance of Reproducibility of Experimental Results Especially in the Domain of Security

Dmytro Petryk[1], Ievgen Kabin[1], Peter Langendörfer[1,2] and Zoya Dyka[1,2]
[1] *IHP – Leibniz-Institut für innovative Mikroelektronik,* Frankfurt (Oder), Germany
[2] *BTU Cottbus-Senftenberg,* Cottbus, Germany
{petryk, kabin, langendoerfer, dyka}@ihp-microelectronics.com

*Abstract* — Security especially in the fields of IoT, industrial automation and critical infrastructure is paramount nowadays and a hot research topic. In order to ensure confidence in research results they need to be reproducible. In the past we reported [18] that in many publications important information such as details about the equipment used are missing. In this paper we report on our own experiments that we run to verify the parameters reported in the datasheets that came along with our experimental equipment. Our results show that there are significant discrepancies between the datasheets and the real world data. These deviations concern accuracy of positions, movements, duration of laser shots etc. In order to improve reproducibility of results we therefore argue on the one hand that research groups verify the data given in datasheets of equipment they use and on the other hand that they provide measurement set-up parameters in globally accepted units such as cm, seconds, etc.

## I. Introduction

The significance of security can hardly be overestimated in the current geo-political situation. It is of utmost importance especially in the IoT, critical infrastructures etc. These systems are to a considerable amount build of embedded systems which are resource constraint and often deployed in areas that allow physical access to those devices. This means that security means need to be researched and realized extremely carefully taking into account physical attacks such as side channel attacks and fault injection attacks.

Unfortunately many papers do not allow reproducibility. In [18] we reported on missing information about experimental set-ups which disregards the rules of good scientific practice and prevents other researches form verifying results reported. But even if all these data are given the reproducibility is often limited as the parameters are vendor specific i.e. the experiments cannot be repeated with equipment from another vendor. In addition we learned in the last years that statements of vendors regarding operating conditions of their equipment such as accuracy, duration energy applied etc. are not fully accurate [13].

Therefore we are reporting in this paper about our experiments and findings in validating parameter settings of our equipment. We hope that this on the one hand encourages other research groups to validate the parameters of their equipment and on the other hand to include accurate parameters in globally accepted units in their papers.

The paper is structured as follows. Section II discusses examples of experiments where documented and real parameters deviate which and in which relying on the documented parameters would lead to inaccurate results. Section III briefly describes optical FI attacks, the equipment/setup for laser FIs sold by Riscure as well as our results of setting parameters evaluation. Section IV concludes this work.

## II. Consequences for Practical Experiments

In this work, we present a practical verification of setup setting parameters on the example of the Diode Laser Station (DLS) from Riscure [1] intended to perform optical Fault Injection (FI) attacks. Many researchers used the laser FI equipment from Riscure in their experiments. A brief search in google scholar using keywords "Riscure" and "laser" resulted in 199 publications[2].

The evaluation of setting parameters (see Section III-C) confirms the fact that parameters given in datasheets may deviate from real ones. Such differences cause unexpected effects to experiments performed, leading to inaccurate ("faulty") results or reduce the reproducibility of experiments.

### A. Laser beam pulse duration

In our recent work, we evaluated influence of laser illumination on static power consumption of NAND cell [13]. To measure static currents we used a high precision Ammeter [17] that is able to measure very low currents. The issue is that to measure very low static currents the analog-to-digital converter of the Ammeter needs at least 10 µs to process its input signal. According to Riscure manual [3] a laser beam pulse up to 100 µs can be generated by the control unit (VC glitcher) for each available laser in the setup. But our measurements show that laser beam pulse duration in the setup is limited to around 1 µs (maximum), see Section III-C. With such a short pulse duration, we would not be able to measure the current reliably, when the pulse duration is set to 100 µs in the Riscure Inspector FI software. Moreover, if we trust the documentation, i.e. if we are sure that the laser pulse duration is 100 µs, and we do not observe the expected increase of the static current, the possible vulnerability of the chips to laser illumination attacks analysing their static power consumption can go unnoticed. But this vulnerability can be used by attackers as a means to compromise critical devices. To be able to perform the measurements we used an FPGA to activate the laser for 100 µs instead of using the Riscure software and VC glitcher. I.e. relying on the datasheet provided we wouldn't have been able to confirm an increase of static power consumption under laser illumination.

In the other case we aimed at manipulating a single memory element during several clock cycles. With the data given in the datasheets it would not have been possible to manipulate, e.g. a flip-flop cell of a shift register working at a clock signal frequency of 10MHz, for more than 10 clock cycles. But this experiment was successful and published in [14] and [15].

---

[1] Riscure is an independent renowned commercial company that performs security certification of semi-conductor products, e.g. embedded systems or smart cards.

[2] The search performed in April 2024. Only data in pdf format was searched.

*B. Laser beam spot size*

In our experiments with radiation-hard shift registers two duplicated transistors that are placed at a distance of 9 µm, have to be influenced simultaneously to manipulate the logic state by illumination. The successful experiment was reported in [14]. We were able to manipulate the logic state of the cell using the single-mode laser with 5× and 20× magnification objectives. Taking into account the 4 µm spot size of the red single-mode laser using a 20× magnification objective as stated in the Riscure datasheet [4], it should be infeasible to manipulate both transistors placed at a distance of 9 µm. Our successful manipulation can be explained by the inconsistency between the laser beam spot size given in the datasheet [4] and the real one. This assumption was confirmed measuring the laser beam spot size. We measured the spot size using reflections (see **Fig. 4**) as well as using a profiler [12] (see TABLE III). **Fig. 1** depicts a part of the attacked cell and laser beam spots using 20× magnification objective and our evaluation.

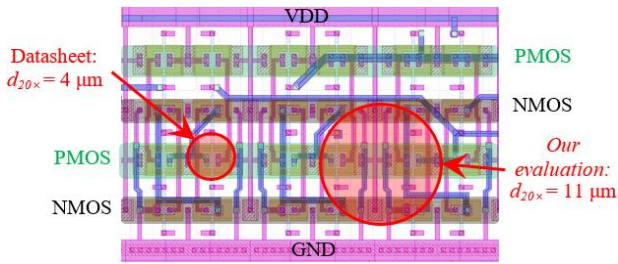

**Fig. 1.** Part of the attacked cell and comparison of laser beam spot sizes taken from the datasheet and experimentally measured.

## III. EVALUATING PARAMETERS USED IN OPTICAL FI ATTACKS

*A. Optical FI attacks*

Optical FI fall under semi-invasive class of attacks and are feasible due to the sensitivity of semi-conductors to light. These attacks are usually realized using lasers, due to their small area of influence and accurate timing, but can also be performed using a conventional camera Flash light [1]. They are frequently used to perturb the internal logic state of memory or logic cells of a device illuminating a small area of the device using a controlled light source. A successfully injected fault can, e.g. cause sensitive/secret data leakage, change execution flow or can be exploited to recover a secret information such cryptographic keys. Details about optical FI attacks as well as an overview of the attacks against different cryptographic implementations, logic cells, volatile and non-volatile memory can be found in [2].

*B. Setup*

We verified setting parameters of a modified 1st generation DLS from Riscure [3]. The DLS consists of: a laser, a control unit for laser (VC glitcher), a DLS body, a microscope camera, magnification objectives and X-Y stage. The setup is equipped with two multi-mode 808 nm and 1064 nm lasers from Riscure [3] and a single-mode 808 nm laser from Alphanov [4]. To control laser beam power and pulse duration the corresponding signals are sent from VC glitcher [6]. Here, the VC glitcher with firmware version 2.3 was used. The setup is controlled by Riscure Inspector FI software [5] installed on a PC, where the end-user does not have access to control signals exchanged. In our evaluations, two versions of the FI software were used: 4.9.1 and 4.12.3. To control the laser beam spot size different long working distance magnification objectives from Mitutoyo are used [7]. To be able to illuminate different parts of a device the device has to be moved relative to the laser beam. For this purpose, the setup includes an X-Y stage [8] that is controlled by a Tango Desktop device [9]. Both are produced by Märzhäuser Wetzlar GmbH & Co [10]. A more detailed description of the setup and the purpose of each device in it can be found in [2].

*C. Evaluation of Setting Parameters*

There are some differences in the device parameters, when comparing the parameters from the Riscure documentation with the parameters from the datasheet of the device manufacturer. E.g. the minimal movement speed of the stage using the Inspector FI software is obviously higher than the movement speed stated in the Märzhäuser Wetzlar GmbH & Co datasheet [9]. To ensure reproducibility of experimental results, the compliance of real setting parameters to the ones stated in [3], [4], [6] and [9] was verified.

We evaluated the following important setting parameters of the DLS:
- Control signals from VC glitcher:
  o Laser beam power;
  o Laser beam pulse duration;
- Laser beam spot size;
- Accuracy of the chip positioning.

*1) Signals from VC glitcher controlling the laser pulse*

The laser beam output power determines the intensity emitted per unit of time. The laser beam pulse duration determines the time for which the laser is active (illuminating). The precise control over these parameters is very important since it may lead to different results ranging from no measurable influence to damaging a device.

According to [3] and [6] the VC glitcher generates signals as follows:
- A linear signal in a range from 0.0 V up to 3.3 V, where 0.0 V corresponds to 0 % and 3.3 V – 100 % laser beam output power. The minimal step is 1 %.
- A binary signal in a range from 0.0 V up to 3.3 V, with a threshold at 2.4 V, i.e. from 0.0 V to 2.4 V corresponds to signal level *low* and from 2.4 V to 3.3 V – signal level *high*. The pulse duration is determined by the duration of a signal in level *high*.

Since we were unable to measure the laser beam power or the laser pulse duration directly, due to the absence of specialized equipment, we experimentally evaluated the control signals sent to the laser. The control signals from the VC glitcher were measured with different sampling rates from 250 MS/s to 10 GS/s using a Teledyne Lecroy WavePro 254HD oscilloscope [16].

*a) Laser beam output power*

According to our measurements, the voltage levels of the signal controlling the laser beam output power differ from those stated in [3] and [6]. The results of our measurements are given in TABLE I.

Each measured value in TABLE I was obtained by averaging 3 measurements.

TABLE I. VOLTAGE LEVELS AT THE VC GLITCHER OUTPUT CONTROLLING LASER BEAM OUTPUT POWER

| Inspector software | Voltage levels at the output of the VC glitcher | | Relative difference δ between documented and measured values |
|---|---|---|---|
| | Datasheet (documented) | Measured | |
| 0 % | 0.0 V | 0.26 V | - |
| 25 % | 0.8 V | 1.04 V | 30 % |
| 50 % | 1.7 V | 1.83 V | 8 % |
| 75 % | 2.5 V | 2.63 V | 5 % |
| 100 % | 3.3 V | 3.44 V | 4 % |

a. Laser beam output power in the Inspector software is set in from 0 % to 100 % with a step of 1 %.

For comparison of the values given in the Riscure datasheet and our measured values we used the relative difference δ between the documented and the measured values:

$$\delta = \frac{|documented\_value - measured\_value|}{documented\_value} \cdot 100\% \quad (1)$$

According to the measurements, the voltage levels differ significantly from the documented values of "power" when set to 25 % or less. Starting from 50 % of "power" and more, the difference is less significant, see TABLE I. Our measurements show that the VC glitcher always sets the voltage to 2.47 V before the first laser shot. Then the voltage is adjusted to the set value, e.g. if 50 % laser beam power was set the voltage decreases from 2.47 V to 1.83 V, see **Fig. 2**.

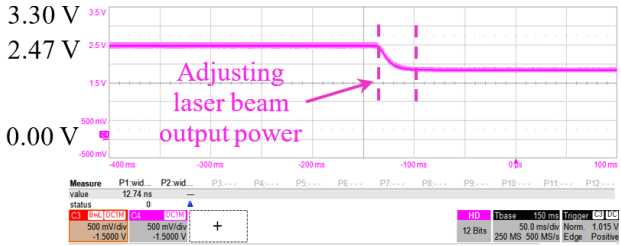

**Fig. 2.** Measured oscilloscope waveform of the "laser beam output power" signal generated at the output of the VC glitcher. The voltage level adjusts to 1.83 V for the first laser shot, i.e. 50 % "power".

For the subsequent shots, the voltage adjusts from the last voltage set to the needed one. Our measurements show that voltage levels can be set with a high accuracy. This fact is highly important for reproducibility of the FI experiments.

*b) Laser beam pulse duration*

According to [6] the VC glitcher is able to generate signals with a duration from 2 ns up to $2^{31}$ ns with a step of 2 ns. The Riscure multi-mode lasers can only be operated in a pulse mode with a duration of pulses from 20 ns up to $10^5$ ns. TABLE II gives the results of pulse duration measurements[3].

Each measured value in TABLE II was obtained by averaging 3 measurements.

Our measurements show that the durations of the control signal at the output of the VC glitcher differ from those given in [6]. Setting the pulse duration in the Inspector FI software to:

- 2 ns or 4 ns results in an about 5 ns long pulse;
- 1000 ns or longer results in a 997 ns long pulse.

TABLE II. DURATION OF SINGNAL AT LEVEL *HIGH* AT THE VC GLITCHER OUTPUT CONTROLLING LASER BEAM PULSE DURATION

| Inspector software | Pulse duration at the VC glitcher output | | Relative difference δ between documented and measured values |
|---|---|---|---|
| | Datasheet (documented) | Measured | |
| 2 ns | 2 ns | 4.6 ns | 130 % |
| 4 ns | 4 ns | 4.7 ns | 18 % |
| 6 ns | 6 ns | 5.8 ns | 3 % |
| 10 ns | 10 ns | 10.0 ns | 0 % |
| 20 ns | 20 ns | 22.6 ns | 13 % |
| 50 ns | 50 ns | 48.4 ns | 3 % |
| 100 ns | 100 ns | 96.6 ns | 3 % |
| 250 ns | 250 ns | 247.4 ns | 1 % |
| 1000 ns | 1000 ns | 997.3 ns | 0 % |
| 1500 ns | 1500 ns | 997.3 ns | 33 % |
| 2000 ns | 2000 ns | 997.3 ns | 50 % |
| $10^4$ ns | $10^4$ ns | 997.3 ns | 90 % |
| $10^5$ ns | $10^5$ ns | 997.3 ns | 99 % |

Nevertheless, the reproducibility of measured values is high. **Fig. 3** shows an oscilloscope waveform of the "laser beam pulse duration" signal generated at the output of the VC glitcher setting different pulse durations in the Inspector FI software.

Please note that we do not exclude the fact that pulses longer than 997 ns can be achievable under some specific operating conditions of the VC glitcher.

*2) Laser beam spot size*

The laser beam spot size determines the area of the illumination. The Riscure documentation does not provide any information on the:

- intensity distribution of the multi-mode lasers;
- exact spot size.

To evaluate the spot sizes we:

- calculated the theoretical spot sizes using the formula of minimal laser beam spot diameter[4] *d* for a given Numerical Aperture (NA) of a magnification objective using the following formula:

$$d = \frac{1.22\lambda}{NA} \quad (2)$$

where $\lambda$ is a wavelength (see more details in [7], [11]).
- captured reflections of laser beam from substrate surface using different magnification objectives[5].
- used a laser beam profiler [12].

Due to the absence of a laser beam profiler in our first experiments we evaluated laser beam spot sizes using reflections of the laser beam from a silicon surface. **Fig. 4** shows the reflections of the laser beam spots from silicone surfaces for the red single-mode and the red multi-mode laser as well as images captured later using a laser beam profiler for the red single-mode laser. To achieve the smallest beam waists the laser beam was focused on the die surface using different magnification objectives. To exclude influence of surface roughness illumination was done at different areas. Our observations show that the shape and the size of laser beam reflection remain unchanged when illuminating different areas. The sizes of the reflections were determined relative to the areas captured by the microscope camera [3]

---

[3] Measurements are done at 1 MHz operating frequency. Details about operating frequency can be found in [2].
[4] The formula is only valid for single-mode laser beams.
[5] Measurements of laser beam spot sizes were performed with the modified DLS.

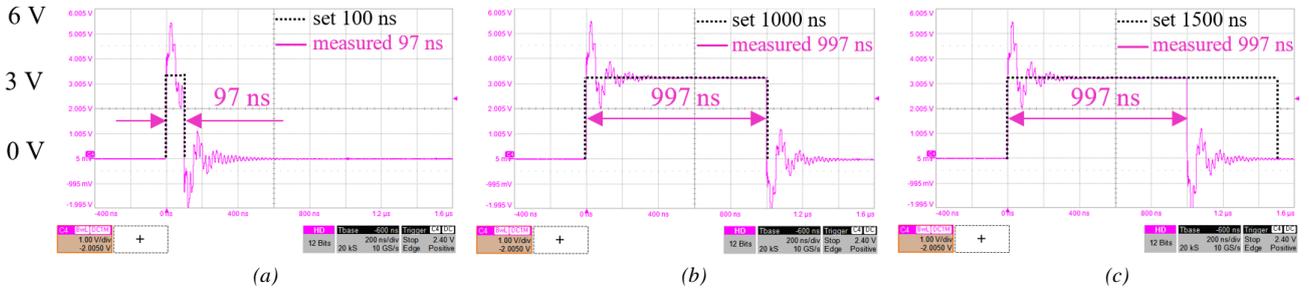

**Fig. 3.** Measured oscilloscope waveform of the "laser beam pulse duration" signal generated at the output of the VC glitcher. The signal durations were set in the Inspector FI software as follows: *(a)* – 100 ns; *(b)* – 1000 ns; *(c)* – 1500 ns.

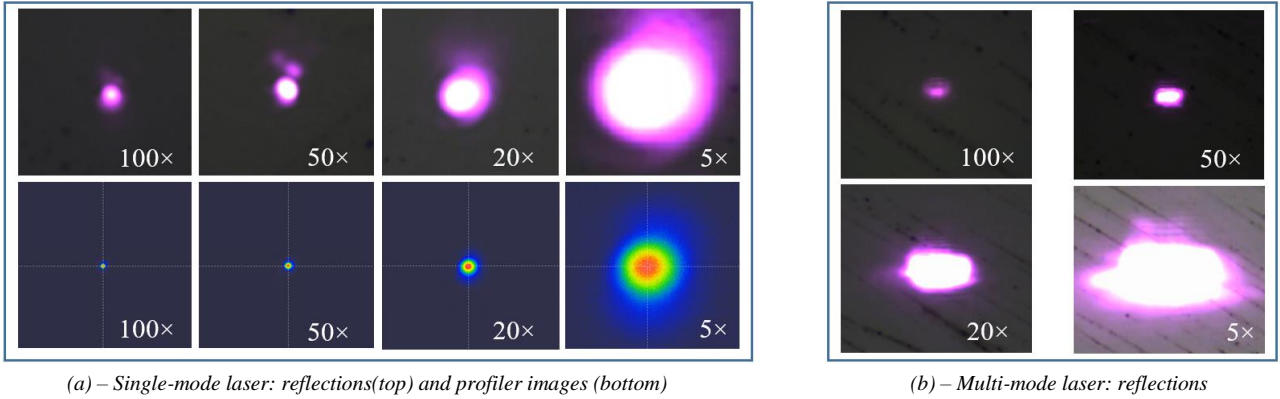

*(a) – Single-mode laser: reflections(top) and profiler images (bottom)*      *(b) – Multi-mode laser: reflections*

**Fig. 4.** Reflections of the laser beam spots from silicone surface for the red single-mode and the red multi-mode laser as well as images captured using a laser beam profiler for the red single-mode laser.

using the knowledge about the number of camera pixels per 1 µm$^2$ for each of the used objectives. We are aware that these measurements do not give very accurate results, especially if the expected laser beam spot size is about $d = 1$ µm.

Using the laser beam profiler we measured the spots of the red single-mode laser in the continuous wave mode at $1/e^2$, i.e. the spot size with 86.5 % intensity. The maximum intensity of the laser beam in this mode is limited to around 30 %, i.e. laser beam spot sizes in pulsed mode are expected to be bigger. TABLE III gives the results of our evaluations.

According to our calculations, evaluations and measurements the laser beam spot sizes differ significantly from the ones given in the datasheets [3] and [4].

*3) Chip positioning*

According to our observations, the minimal movement speed of the X-Y stage using the Riscure Inspector FI software is higher than the speed stated in [9]. The movement speed is expected to influence the accuracy of the step sizes due to inertia. Small inaccuracies caused by inertia can lead to significant differences in the theoretically calculated and the practically reached coordinates for laser illumination. Precise targeting is also important if areas sensitive to laser have to be determined. Hence, we evaluated experimentally the minimum step size *s* and step accuracy $\Delta_s$. The parameters, i.e. *s* and $\Delta_s$, define the distance between 2 neighboring stops of an X-Y stage, which is $s \pm \Delta_s$ and can depend on the speed of the X-Y stage movements. The influence of the speed on the step accuracy Δs, and – consequently – on the chip positioning accuracy in the measurement setup was not found neither in the Tango datasheet [9] nor in the Riscure manual [3].

TABLE III. RESULTS OF LASER BEAM SPOT SIZE EVALUATIONS

| Magnification objective | Laser beam spot size | | | | | |
|---|---|---|---|---|---|---|
| | Single-mode laser | | | | Multi-mode laser | |
| | Calculated | Datasheet[a] [4] | Reflection measure- ments[a] | Profiler measure- ments | Manual [3] | Reflection measure- ments |
| | $d^c$, µm | $d$, µm | | | Spot size, µm$^2$ | |
| 100× | 1.97 | 1 | 1.45 | 1.62 | 3.0×0.8[b] | 5.85×3.30 |
| 50× | 2.35 | 1.5 | 3.20 | 2.59 | 6×1.4 | 8.60×5.20 |
| 20× | 2.46 | 4 | 11.00 | 5.43 | 15×3.5 | 29.50×17.50 |
| 5× | 7.04 | 15 | 45.00 | 21.91 | 60×14 | 150.00×74.00 |

a. Laser beam spot sizes given in the Riscure manual [3] correspond to spot sizes where 80 % of the energy is concentrated values given in [4] are expected to be measured in the same way.
b. The diameter of the first Airy disk, i.e. the diameter between the points where laser beam intensity reduces to 0 %.
c. Expected laser beam spot size that is derived from laser beam spot sizes given for 50×, 20× and 5× magnification objectives.

According to [8], the stage is controlled by two spindles (to move along *x* and *y* axis) with a 1 mm pitch. The stage's minimal movement speed amounts to 10$^{-6}$ revolutions per second and the minimal step size to 0.01 µm [8]. We assume that 0.01 µm step size is only achievable at the minimal possible speed of the Tango Desktop unit. This means that about 12 days (10$^6$ seconds) are required to move the stage by 1 mm. Subsequently, the minimal linear speed of the stage is 1 mm/10$^6$ seconds=10$^{-9}$ m/s=1 nm/s. The minimal speed that is possible to set in the Inspector FI software using the slider is significantly higher than 1 nm/s. According to our observations to travel 100 µm distance along *x* axis takes about 1 s, if setting the step size to 100 µm and selecting the minimal possible speed in the Inspector FI software[6].

---

[6] Five measurements were performed to evaluate the minimal movement speed of the stage. We are aware about the inaccuracy of time measurements. The speed calculated using this inaccurate time is necessary only to demonstrate that it is much higher than the minimal speed declared in the datasheet.

This implies that using the Riscure software the minimum speed of the stage is about 100 µm/s that is $10^5$ times higher than the calculated speed of 1 nm/s.

Next, we evaluated the minimal step size at the minimal speed of 100 µm/s. The different path lengths and different number of the steps per path were set. Each single step can be observed by a short stop of the X-Y stage using the microscope camera. We measured the distance that the X-Y stage traveled related to the distance in the layout. Due to limited camera resolution we were not able to measure the length of each step precisely. Hence, to determine the minimal step size we calculated the number of "stops" from the start to the end point. Please note that in the Riscure software only the distance between 2 points on the chip and the number of the stops between these 2 points can be set, i.e. not the step size itself. We set the travel distance to 10 µm and experimented with the number of stops. We observed that only setting the number of stops up to 40 stops (for 10 µm distance) results in stable movement of the X-Y stage. Setting the number of stops larger than 40 did not result in repeatable and stable movements of the X-Y stage. Thus, based on our observations we determined the *minimal reasonable step size* of 0.25 µm that is 5 times bigger than 0.05 µm step size stated in [3].

## IV. Conclusion

Security especially in the fields of IoT, industrial automation and critical infrastructure is paramount nowadays. Research in this field can and has to provide innovative and reliable solutions. In order to ensure that these solutions can be applied by industry they need to achieve a high credibility. Credibility essentially requires reproducibility of the results achieved. In the past we reported [18] that in many publications important information such as details about the equipment used are missing. Unfortunately even if these parameters are reported reproducibility is not really ensured as often parameters are vendor specific, i.e. research groups using equipment from another vendor cannot repeat the experiments with a sufficient accuracy. In order to solve this issue we argue that parameters such as positions, movements, durations, voltages etc. are given in globally accepted units e.g. cm, seconds, V, joule etc. In this paper we have presented experiments that clearly show that relying on datasheets of equipment vendors may significantly harm reproducibility as there are significant deviations between e.g. the accuracy of positions according to the datasheet and what we could experimentally verify. So, we recommend that all research groups start double checking the parameters of the equipment they use for experiments and provide the settings used in their experiments in vendor independent units.